# Correlation Properties of Signal at Mobile Receiver for Different Propagation Environments

Cezary Ziółkowski, Jan M. Kelner, Leszek Nowosielski
Institute of Telecommunications, Faculty of Electronics
Military University of Technology
Warsaw, Poland
{cezary.ziolkowski, jan.kelner, leszek.nowosielski}@wat.edu.pl

*Abstract*—**An issue of the parameter selection in various branches of a multi-antenna receiver system determines its effectiveness. A significant effect on these parameters are correlation properties of received signals. In this paper, the assessment of the signal correlation properties for different environmental conditions is presented. The obtained results showed that depending on the receiver speed, the adaptive selection of the delays in the different RAKE receiver branches provide minimization of the correlation between the signals. Particularly low levels of the signal correlation could be obtained in complex propagation environments such as urban and bad urban.**

## I. Introduction

The impact of the environment on the autocorrelation function (ACF) properties is important for the development of multi-antenna systems and processing methods of the received signals. For individual branches of the system, the evaluation of ACF properties allows to determine the parameters that provide minimize the correlation of the received signals. In RAKE receiver [1][2][3][4], knowing ACF properties is the basis for choosing the values of the coefficients and delays of each branch of the tapped delay line. Superposition of all signals from line ensures minimization of frequency selective fading. This mean that ACF properties play a large role in minimizing intersymbol interference and, as a consequence, they significantly affect the capacity of the entire system.

The angular spread of power is one of the main factors that determine ACF properties of the received signal. Hence, the assessment of these properties is determined by selection of the appropriate angular spread model that provides mapping the effects of actual propagation phenomena. The basic function that defines a statistical spread model is the probability density function (PDF) of angle of arrival (AOA). In this set of models, we can distinguish analytical (geometrical) and empirical models. The analytical models are based on the geometrical structures that described the spatial location of the scattering elements. The empirical models whereas use standard PDFs. In this case, the criterion for selection of model parameters is to minimize the approximation error with respect to measurement data. The analytical models characterized by considerable complexity of the analytical description compared to empirical models. In addition, in [5], the comparison of the results of approximation shows that the empirical models provide a smaller approximation error of measurement data with respect to the analytical models.

Influence of the propagation environment on the ACF properties is the main purpose of this paper. The assessment of this impact is based on the selected PDF of AOA empirical model that minimizes the approximation error of data from several measurement campaigns. The modified Laplacian distribution is one of the fundamental empirical models that is used to describe PDF of AOA. The accuracy of measurement data mapping for this model is presented in [6][7], and a comparative assessment with other models is contained in [8]. The presented results are the reason for choosing this model to assess ACF properties of the received signal by mobile receiver in different propagation conditions. Propagation environment type is defined on the basis of the rms delay spread (DS), $\sigma_\tau$. Adaptation of the modified Laplacian uses the relationship between the parameters of this model and DS. Dispersion of AOA versus DS allows the assessment of the impact of propagation environment on AFC properties that among other things, determine the efficacy of using multi-antenna systems. A method of determining the empirical model parameters as a function of DS is shown based on measurement data in [9].

The remainder of the paper is organized as follows. In Section II, we present the theoretical basis for the determination of ACF analytical form of received signal by a moving receiver. This section also shows the PDF of AOA and an adapting method its parameters to the type of propagation environment. The calculation results of averaging procedure for the different environmental parameters are presented in Section III. From the obtained results, final conclusions are contained in Section IV.

## II. Normalized Correlation Function and Angle Distribution

The fundamental model that defines ACF of the received signal by the mobile receiver is a two-dimensional isotropic

Clark model. For this model, the analytic form of the low-pas received signal, $x(t)$, for the unmodulated carrier is [10]

$$x(t) = \sum_{n=1}^{N} a_n \exp\left(2\pi i f_{Dm} t \cos \varphi_n + i \gamma_n\right) \quad (1)$$

where $N$ is the number of propagation paths, $a_n$ represents the signal amplitude of the $n$th path, $f_{Dm}$ is the maximum frequency of the Doppler shift, $\varphi_n$ means angle between the $n$th path and velocity vector of the receiver, $\gamma_n$ is the random phase of signal component from the $n$th path.

In practice, all parameters of $x(t)$ are statistically independent and $\gamma_n$ values are equally likely. It is the premise that PDF of each $\gamma_n$ $(n=1,2,...,N)$ can be assumed to be uniformly distributed. By using these assumptions, the ACF, $R(\tau)$, for time interval that conditions stationary properties of $x(t)$, is

$$
\begin{aligned}
R(\tau) &= \mathrm{E}\left\{x(t) x^*(t-\tau)\right\} \\
&= \mathrm{E}\left\{
\begin{array}{l}
\sum_{k=1}^{N} a_k \exp\left(2\pi i f_{Dm} t \cos \varphi_k + i \gamma_k\right) \\
\times \sum_{l=1}^{N} a_l \exp\left(-2\pi i f_{Dm}(t-\tau) \cos \varphi_l - i \gamma_l\right)
\end{array}
\right\} \\
&= \mathrm{E}\left\{\sum_{n=1}^{N} a_n^2 \exp\left(2\pi i f_{Dm} \tau \cos \varphi_n\right)\right\} \\
&= \sum_{n=1}^{N} \mathrm{E}\left\{a_n^2\right\} \mathrm{E}\left\{\exp\left(2\pi i f_{Dm} \tau \cos \varphi_n\right)\right\}
\end{aligned}
\quad (2)
$$

where $\mathrm{E}\{\cdot\}$ is an averaging operator after sets of $a_n$ and $\varphi_n$ values.

The statistical properties of each AOA are the same for all propagation paths. Thus, we have

$$R(\tau) = \mathrm{E}\left\{\exp\left(2\pi i f_{Dm} \tau \cos \varphi\right)\right\} \cdot \sum_{n=1}^{N} \mathrm{E}\left\{a_n^2\right\} \quad (3)$$

where $\sum_{n=1}^{N} \mathrm{E}\left\{a_n^2\right\}$ is the average power of received signal.

Hence, the normalized ACF, $r(\tau)$ has a form

$$
\begin{aligned}
r(\tau) = R(\tau)/R(0) &= \mathrm{E}\left\{\exp\left(2\pi i f_{Dm} \tau \cos \varphi\right)\right\} \\
&= r_I(\tau) + i r_Q(\tau)
\end{aligned}
\quad (4)
$$

where $r_I(\tau) = \mathrm{E}\left\{\cos\left(2\pi f_{Dm} \tau \cos \varphi\right)\right\}$ and $r_Q(\tau) = \mathrm{E}\left\{\sin\left(2\pi f_{Dm} \tau \cos \varphi\right)\right\}$ are the in-phase and quadrature components of $r(\tau)$.

For the Clark model, PDF of AOA, $f(\varphi)$, is described by the uniformly distributed. In this case, we have

$$r(\tau') = \mathrm{J}_0(2\pi \tau') \quad (5)$$

where $\mathrm{J}_0(\cdot)$ is the zero-order Bessel function of the first kind and $\tau' = f_{Dm}\tau$ is a normalized time.

However, a number of measurement results executed in real environment significantly different from the uniform PDF. This fact is shown in many publications, for example [6]. In this paper, our attention focuses on the empirical model $f(\varphi)$ that provides a simple analytical description and minimizes approximation errors of measurement data. In [8], the comparative analysis of empirical models shows that the modified Laplacian gives the smallest approximation error for different propagation scenarios. Therefore, in the paper, this PDF is used to analyze the signal ACF. Modified Laplacian has a form

$$f(\varphi) = C(\sigma_\tau) \frac{\lambda(\sigma_\tau)}{2} \exp\left(-\lambda(\sigma_\tau)|\varphi|\right) \quad \text{for} \quad \varphi \in \langle -\pi, \pi) \quad (6)$$

where $C(\sigma_\tau) = \left(1 - \exp\left(-\lambda(\sigma_\tau)\pi\right)\right)^{-1}$ is a normalized factor such that $\int_{-\pi}^{\pi} f(\varphi)\mathrm{d}\varphi = 1$.

Adaptation of $f(\varphi)$ to the type of propagation environment consists in appropriately adjusting the PDF parameters. In this procedure, a statistical relationship between DS and the rms angle spread (AS), $\sigma_\varphi$, is used [6]. Details of adaptation method are shown in [9]. Considering the results of several measurement campaigns, the relationship between $\lambda$ and $\sigma_\tau$ is approximated by [9]

$$\lambda(\sigma_\tau) = \left(9.44 \cdot \sigma_\tau + 0.40\right)^{-1} \quad (7)$$

Considering (4) and (6), the analytical forms of the in-phase and quadrature components of ACF, we can express as

$$
\begin{aligned}
r_I(\tau', \sigma_\tau) = &\, C(\sigma_\tau) \frac{\lambda(\sigma_\tau)}{2} \\
&\times \int_{-\pi}^{\pi} \cos\left(2\pi \tau' \cos \varphi\right) \exp\left(-\lambda(\sigma_\tau)|\varphi|\right) \mathrm{d}\varphi
\end{aligned}
\quad (8)
$$

$$
\begin{aligned}
r_Q(\tau', \sigma_\tau) = &\, C(\sigma_\tau) \frac{\lambda(\sigma_\tau)}{2} \\
&\times \int_{-\pi}^{\pi} \sin\left(2\pi \tau' \cos \varphi\right) \exp\left(-\lambda(\sigma_\tau)|\varphi|\right) \mathrm{d}\varphi
\end{aligned}
\quad (9)
$$

Formulas (8) and (9) are the basis to determine the module of ACF, $r(\tau', \sigma_\tau)$, which enables the ACF assessment of the received signal in different environmental conditions.

## III. NUMERICAL CALCULATION

In the paper, we use numerical calculation for analysis of signal ACF. Such an approach to analyzed problem results from the complexity of analytic expressions (8) and (9). Numerical calculation of integrals are made for three different values of $\sigma_\tau$ (0.1 µs, 0.2 µs, 1.0 µs) that define the propagation environments of rural area and typical urban types. The obtained results of $r_I(\tau', \sigma_\tau)$, $r_Q(\tau', \sigma_\tau)$, and $|r(\tau', \sigma_\tau)|$ are presented in Figs. 1-3.

As shown in graphs, the *strong* correlation of the signal occurs for rural area, and $|r(\tau', \sigma_\tau)|$ graph is converges to the Clark model with the increased complexity of the propagation environment. This means that the application of the multi-antenna system in rural area gives a significantly smaller improvement in the quality and channel capacity compared to an environment that has a large $\sigma_\tau$ value. The graph of $|r(\tau', \sigma_\tau)|$ shows that in individual branches, the selection of the delays minimizes correlate the received signals in the multi-antenna system. The minimum value of ACF that can be obtained for the first local minimum ($\tau' = \tau'_{min}$) are shown in Fig. 4 for different propagation environments (versus $\sigma_\tau$). In addition, the maximum values of $|r(\tau', \sigma_\tau)|$ (the first local maximum $\tau' = \tau'_{min}$) are presented in this figure.

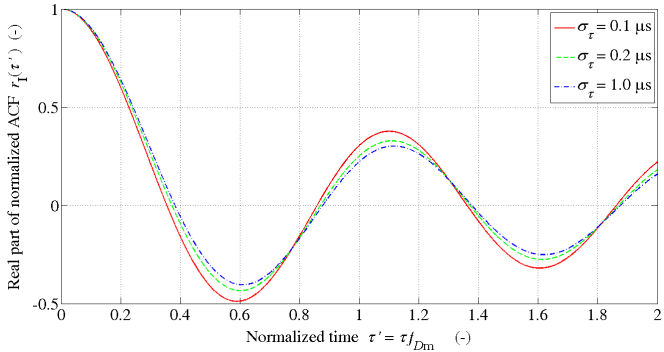

Fig. 1.  In-phase component of ACF for different environment types.

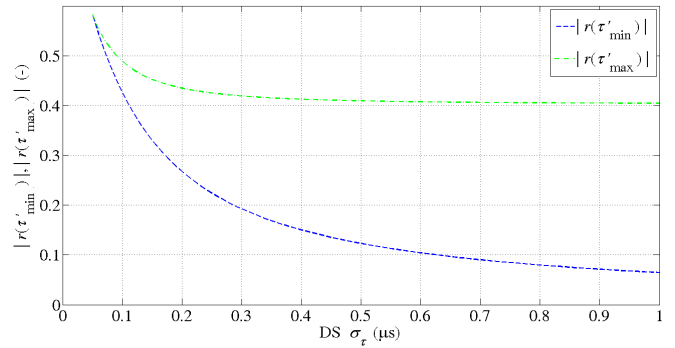

Fig. 4.  First local minimum and maximum of $|r(\tau', \sigma_\tau)|$ for different environment types.

The figure indicates that for the higher the complexity of propagation environment, the ACF value can be a smaller. Comparing graphs shows that the use of procedure for delaying the signals in individual branches can provide the efficiency increase of signal reception.

A measure

$$\Delta r(\sigma_\tau) = |r(\tau'_{max}, \sigma_\tau)| - |r(\tau'_{min}, \sigma_\tau)| \qquad (10)$$

that gives you the opportunity to evaluate the use effectiveness of the first local minimum of the module of ACF in the branches receiver, is shown in Fig. 5.

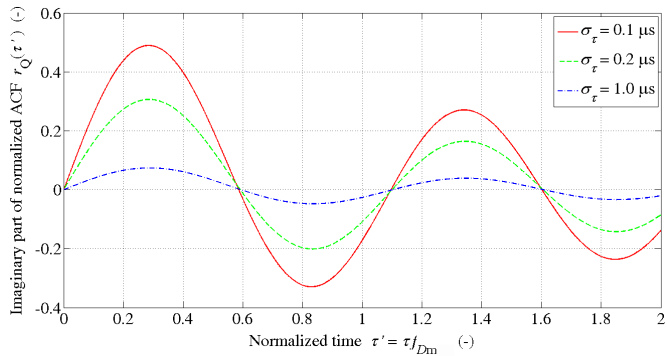

Fig. 2.  Quadrature component of ACF for different environment types.

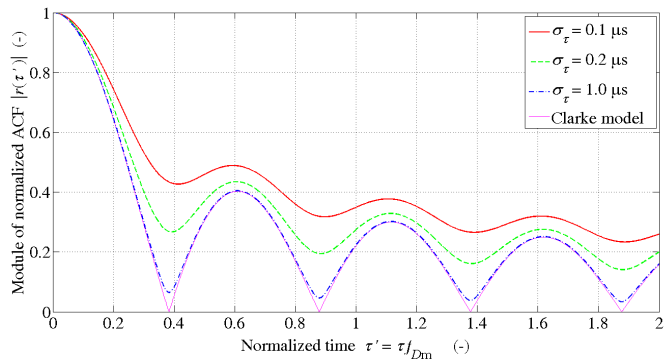

Fig. 3.  Module of ACF for different environment types.

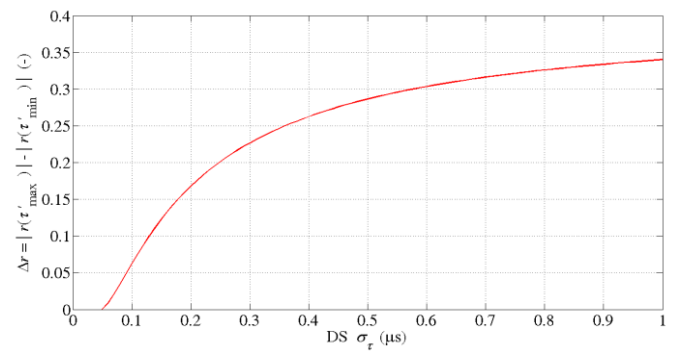

Fig. 5.  Difference between first local maximum and minimum of ACF for different environment types.

The graph of $\Delta r(\sigma_\tau)$ shows that the selection of the signal delays in various branches of the receiver could provide several times reduction of the signal correlation for complex propagation environments such as urban and bad urban areas.

## IV. CONCLUSION

The issue of the parameter selection in the various branches of the multi-antenna system and processing method determines their effectiveness. A significant effect on these parameters are the correlation properties of the received signals. In this paper, the assessment of signal correlation properties for different environmental conditions is presented. The obtained results showed that depending on the receiver speed, the adaptive selection of the delays in the different RAKE receiver branches provide minimization of the correlation between the signals. This mean that ACF properties play a large role in minimizing intersymbol interference and, as a consequence, they significantly affect the capacity of the entire system. Particularly low levels of the signal correlation could be obtained in complex propagation environments such as urban and bad urban.